\documentclass[11pt,numbers,sort&compress]{elsarticle}   	
\usepackage{amssymb}

\usepackage{lineno,hyperref,geometry,amsmath,appendix,color}
\modulolinenumbers[5]
\usepackage{bbold}
\newcommand{\gdualn}[1]{\overset{\:{}^{{}^{\boldsymbol{\neg}}}}{\smash[t]{#1}}} 
\def\0{\mbox{\boldmath$\displaystyle\mathbb{O}$}}

\def\R{\mbox{\boldmath$\displaystyle\mathbb{R}$}}

\def\x{\mbox{\boldmath$\displaystyle\boldsymbol{x}$}}

\def\0{\mathbb{0}}
\def\I{\openone}
\def\gb{\mbox{\boldmath$\displaystyle\boldsymbol{\gamma}$}}
\def\s{\mbox{\boldmath$\displaystyle\boldsymbol{\sigma}$}}
\def\p{\mbox{\boldmath$\displaystyle\boldsymbol{p}$}}
\def\q{\mbox{\boldmath$\displaystyle\boldsymbol{q}$}}

\def\openone{\mathbb I}
\def\pii{\mbox{\boldmath$\displaystyle\boldsymbol{\pi}$}}
\def\a{\mbox{\boldmath$\displaystyle\boldsymbol{a}$}}

\journal{EPL}

\begin{document}

\begin{frontmatter}


\title{Spin-half bosons with mass dimension three half:  towards a resolution of the cosmological constant problem}



\author[mymainaddress]{Dharam Vir Ahluwalia\corref{mycorrespondingauthor}}
\cortext[mycorrespondingauthor]{Corresponding author}
\ead{dharam.v.ahluwalia@gmail.com}


\address[mymainaddress]{
Department of Physics,
Indian Institute of Technology Guwahati,
Guwahati  781 039, Assam, India.}

\begin{abstract}

 We have constructed a very different type of particle than any presently known. It is a boson and resides in the $(1/2,0)\oplus(0,1/2)$ representation space. The associated local field has mass dimension  three half. These new bosons can only be created or destroyed in pairs. When paired with a fermion the total zero point energy of the boson-fermion system identically vanishes. This communication presents the new quantum field theoretic construct. 
\end{abstract}

\end{frontmatter}
\vspace{21pt}

\noindent
\section{The basic assertion, and the departure}


The usual textbook treatments that assign statistics, fermionic or bosonic, to a field hinge on demanding that, modulo the zero point contribution, the field energy be positive definite. Steven Weinberg, on the other hand, argues that for the S-matrix to be Lorentz invariant the Hamiltonian density at two spacelike events must commute.  In the ensuing  development of theory of quantum field  he arrives at the celebrated results on the assignment of spin and statistics~\cite{PhysRev.133.B1318,Weinberg:1995mt}. 
Both approaches carry their own strengths. Here, we provide a counter example to the theorem by proceeding in a manner that combines both the approaches.
For a comparative study of the two approaches we refer the reader to~\cite{MASSIMI2003621}.

Our construct evades  the canonical wisdom on the relation between spin and statistics~\cite{MASSIMI2003621}. All treatments of the theorem, with an important exception of already cited papers of Weinberg, fail to notice that the expansion coefficients in a field do not have a phase freedom, and that the Lagrangian density is not to be assumed but derived by first introducing a quantum field, and then calculating the vacuum expectation value of the time ordered product of the field and its adjoint. The latter requires a definition of the duals for the expansion coefficients. In the work on mass dimension one fermions these observations were explicitly implemented; as the work evolved from a non-local theory to a local theory~\cite{Ahluwalia:2004sz,Ahluwalia:2004ab,Ahluwalia:2009rh,Ahluwalia:2016rwl,Ahluwalia:2015vea,Rogerio:2016mxi,Ahluwalia:2019etz}.


In the fermionic Dirac field, our reader would recall that  the `particle' expansion coefficients  have a positive norm, while the `antiparticles' expansion coefficients  have a negative norm. These signs coupled with the anticommuting  annihilation and creation operators result in a, except for the sign of the zero point energy,   positive definite field energy. The mass dimensionality of three half can similarly  be traced to the spin sums in which the Dirac dual plays a crucial role. 

Parallel with the theory of mass dimension one fermions,  all the four expansion coefficients of the field we present here 
have a vanishing  norm; that is: if we work with the Dirac dual. We are thus forced to introduce a new dual. Under this new dual \textit{all four} expansion coefficients have a positive Lorentz invariant norm. 
We find that, (a) the field and its adjoint commute for spacelike separation, and (b)  the field energy is positive definite. 

These results arise from an interplay between the norms and spins sums of the spinors, and commuting  annihilation and creation operators.
The Lorentz covariant spin sums and invariant norms alter in such a way that we obtain not only a bosonic 
field but also a local field of mass dimension three half.


\subsection{Outline} 

The next section opens with an observation that served as a seed, and resulted in the reported construct.  In section~\ref{Sec:ExpansionCoefficients} we provide the expansion coefficient for the new field. The new dual, orthonormality relations and spin sums are given in 
section~\ref{Sec:Dual}. Section~\ref{Sec:CliffordAlgebra} presents what we call `a bosonic representation of Clifford algebra.' 
 With this preparation, the new field and its adjoint is introduced  in section~\ref{Sec:quantumfield}. The bosonic nature of the field is established in sub-section~\ref{sec:bosonic} followed by, in sub-section~\ref{Sec:FDPropagator}, the Feynman-Dyson propagator. A very brief sub-section~\ref{Sec:locality} establishes the  new field to be local. That the field energy is bounded from below with a positive definite zero point energy is shown in sub-section~\ref{Sec:FieldEnergy}. The last section presents a few concluding remarks, followed by  an Appendix that collects together some parenthetic information.

\section{The opening observation} 

The narrative that follows is neither fully compatible with history nor with the various textbook presentations. It departs from Weinberg's formalism without abandoning his insights.

Let's go back to the period around 1928 and in that mind set note that the  Dirac equation arose in taking the square root of the dispersion 
relation~\cite{Dirac:1928hu,BrauerWeyl:1935}\footnote{Our notation is standard, the spacetime metric has the signature $\{+,-,-,-\}$, and the basis Weyl~\cite{Ryder:1985wq}.\label{fn:f}}
$$p_\mu p^\mu = m^2 \label{eq:dr}$$
The square root of the left hand side was found to be $\gamma_\mu p^\mu$, where the $\gamma_\mu$ are the $4\times 4$ matrices of the Dirac framework. The argument naturally leads to 
$(\gamma_\mu p^\mu \pm m \I)\psi(\p) =0$, the Dirac equation in momentum space with the $\gamma_\mu$ satisfying the Clifford algebra: $\{\gamma_\mu,\gamma_\nu\} = 2 \eta_{\mu\nu}$.
Its solutions, after incorporating certain locality phases~\cite{Ahluwalia:2016jwz}, later became expansion coefficients for all the fermionic matter fields of the standard model~\cite{Weinberg:1995mt}.

Modulo the Majorana observation of 1937~\cite{Majorana:1937vz}, there is a general consensus that the Dirac field is a unique spin one half field that is consistent with Lorentz symmetries, principles of quantum mechanics, and the cluster decomposition principle.
The uniqueness, with the exception of Weinberg formalism,\footnote{Our departure from Weinberg shall be apparent as we proceed -- it lies in a new dual and the associated adjoint.}  hinges on the implicit assumption that the square root of a $4\times4$ identity matrix $\I$ multiplying the $m^2$  on the right hand side of~(\ref{eq:dr}) is, apart from a mutiplicative factor of $\pm 1$, $\I$ itself -- uniquely.  The recent emergence of the new spin one-half fermions with mass dimension one provides a strong reason that other roots of $\I$ may lead to new spin one half matter fields, and these may serve the dark matter sector or at the least  provide us with a complete set of particle content consistent with basic principles of quantum mechanics and the symmetries of special relativity~\cite{Ahluwalia:2019etz}.

With this background we recall the well known linearly independent square roots of $4\times 4$ identity matrix in the $(1/2,0)\oplus(0,1/2)$ representation space~\cite[page 71]{Schweber:1961zz}
\begin{align}
&\I \nonumber\\
&i\gamma_1 \quad  i \gamma_2\quad i\gamma_3 \quad\gamma_0 \nonumber\\
& i \gamma_2\gamma_3\quad i\gamma_3\gamma_1 \quad i \gamma_1\gamma_2\quad\gamma_0\gamma_1\quad\gamma_0\gamma_2\quad\gamma_0
\gamma_3\nonumber\\
&i\gamma_0\gamma_2\gamma_3\quad
i\gamma_0\gamma_1\gamma_3\quad
i\gamma_0\gamma_1\gamma_2\quad
\gamma_1\gamma_2\gamma_3\nonumber\\
&i\gamma_0\gamma_1\gamma_2\gamma_3
\end{align}
We denote these by  $\Omega_\ell$, $\ell=1,2\cdots 16$:  $\Omega_1$ being the first entry in the above array and $\Omega_{16}$ being the last -- that is, $\ell$ assignment is in consecutive order. As noted in footnote~\ref{fn:f}, we shall adopt the Weyl basis.
These observations are taken from a related work of the author and are reproduced here for the sake of establishing a context.

Unlike the Dirac root $\I$,  none of the remaining roots commute with $\gamma_\mu p^\mu$
\begin{equation}
\left[\Omega_\ell,\gamma_\mu p^\mu\right]  \ne 0 \quad \ell = 2,3, \cdots 16\label{eq:nc}
\end{equation}
In addition, the $\Omega_\ell$ either commute or anti-commute with each other
\begin{equation}
\left[\Omega_\ell,\Omega_{\ell^\prime}\right]_\pm = 0\label{eq:pm}
\end{equation}
Where the symbol $\pm$ denotes commutator, for the minus sign,  or anticommutator, for the plus sign. Some additional details are given in Table~\ref{table1}.

Dirac equation, and the Dirac quantum field, follows from using $\Omega_1=\I$ as the trivial square root of $\I$ provided
the four-fold degenerate eigenspinors of $\I$
\begin{equation}
\nu_1 = \left(
\begin{array}{c}
1 \\
0\\0\\0
\end{array}
\right),\quad\nu_2 = \left(
\begin{array}{c}
0 \\
1\\0\\0
\end{array}
\right),\quad
\nu_3 = \left(
\begin{array}{c}
0 \\
0\\1\\0
\end{array}
\right),\quad
\nu_4 = \left(
\begin{array}{c}
0 \\
0\\0\\1
\end{array}
\right)
\end{equation}
are superimposed to obtain eigenspinors of the 
Dirac root\footnote{See~\cite[Section 5.3]{Ahluwalia:2019etz} on how one obtains the $m^{-1} \gamma_\mu p^\mu$ as the parity operator in the $(1/2,0)\oplus(0,1/2)$ representation space without reference to Dirac equation.}   $\gamma_\mu p^\mu$ at rest ($\p=0$)\footnote{The below enumerated properties also remain true after the application of the $(1/2,0)\oplus(0,1/2)$ boost.}

\begin{equation}
u_+(0)=\sqrt{\frac{m}{2}}\Big(\nu_1(0) + \nu_3(0)\Big),\quad
u_-(0)=\sqrt{\frac{m}{2}}\Big(\nu_2(0) + \nu_4(0)\Big) 
\end{equation}
and 
\begin{equation}
v_+(0)= \sqrt{\frac{m}{2}}\Big(\nu_2(0) - \nu_4(0)\Big),\quad
v_-(0)= -
\sqrt{\frac{m}{2}}\Big(\nu_1(0) - \nu_3(0)\Big) \label{eq:v}
\end{equation}
The choice of phases made above follows from arguments 
presented in references~\cite{Weinberg:1995mt,Ahluwalia:2019etz,Ahluwalia:2016jwz}.
These are independent of 
$\p$ -- and hence can be interpreted as `rest spinors.'\footnote{We are working in the Weyl basis. So, the resemblance of the $\lambda_i(0)$ with the Dirac representation rest spinors is accidental.} Following a procedure that we now embark upon one can construct a local quantum field, that of Dirac, which is fermionic and carries mass dimension three half -- the general procedure can be found in~\cite{Ahluwalia:2019etz}.

With these observations we now embark on a concrete example culminating in the new bosonic field of spin one half.

\section{The expansion coefficients}  \label{Sec:ExpansionCoefficients}

We arrived at the abstracted results while exploring $\Omega_4$.
Its eigenspinors, up to constant multiplicative factors of the type $\alpha \times e^{i \beta}$ with $\alpha,\beta\in\mathbb{R}$, are
\begin{equation}
\mu_1 = \left(
\begin{array}{c}
 0 \\
 i \\
 0\\
 1 \\
\end{array}
\right),\quad
\mu_2 =  \left(
\begin{array}{c}
 i \\
 0 \\
 1 \\
 0 \\
\end{array}
\right),\quad
\mu_3=\left(
\begin{array}{c}
 0 \\
 -i \\
 0 \\
 1 \\
\end{array}
\right),\quad
\mu_4=\left(
\begin{array}{c}
 -i  \\
 0 \\
 1 \\
 0 \\
\end{array}
\right)\label{eq:lambda}
\end{equation}
with eigenvalues, $-1,+1,+1,-1$ respectively. The theory is built upon
the following linear superposition\footnote{Remarks on this step shall be provided in due course. They run parallel to those that introduce the Dirac expansion coefficients~(\ref{eq:v}).  }
 of the $\mu_i $
\begin{equation}
\lambda_1(0) = \sqrt\frac{m}{2}\Big(\mu_1+ i \mu_2\Big),\quad
\lambda_2(0) = \sqrt\frac{m}{2}\Big(\mu_1- i \mu_2\Big) \label{eq:l12}
\end{equation}
and
\begin{equation}
\lambda_3(0) = \sqrt\frac{m}{2}\Big(\mu_3+ i \mu_4\Big),\quad
\lambda_4(0) = \sqrt\frac{m}{2}\Big(\mu_3 - i \mu_4\Big) \label{eq:l34}
\end{equation}
which we treat as eigenspinors at rest.
The action of the $(1/2,0)\oplus(0,1/2)$  boost operator
\begin{equation}
\kappa = \sqrt{\frac{E+m}{2 m}}\left(\begin{array}{cc} 
\I + \frac{\s\cdot\p}{E+m} & \0 \\
\0 &  \I- \frac{\s\cdot\p}{E+m}
\end{array}\right)
\end{equation}
on these spinors  yields the four spinors for an arbitrary momentum \begin{equation}
\lambda_i(\p) = \kappa\lambda_i(0)
\end{equation} 
In section~\ref{Sec:quantumfield} we will define a new quantum field with $\lambda_i(\p)$ as its expansion coefficients.

\section{The dual of $\lambda_i(\p)$, orthonormality relations,  and spin sums}\label{Sec:Dual}

As in the case for eigenspinors of the charge conjugation operator~\cite{Ahluwalia:2019etz} -- Eigenspinoren des Ladungskonjugationsoperators (Elko) -- 
 here too we find that under the Dirac dual, each of the  $\lambda_i(\p)$, $i=1,2,3,4$,  has null norm. Based on the insights gained in author's monograph on mass dimension one fermions~\cite{Ahluwalia:2019etz}
we define a new dual\footnote{I wish to thank Sebastian Horvath, Daniel Grumiller, Cheng-Yang Lee, Dimitri Schritt for various contributions to this development.}
\begin{equation}
 \gdualn{\lambda}_i (\p)  =\big[\mathcal{P} \lambda_i(\p)\big]^\dagger\gamma_0
\label{eq:pod}
\end{equation}
for it allows us to construct an internally consistent theory and provides Lorentz invariant norms. In the above definition $\mathcal{P} =m^{-1} \gamma_\mu p^\mu$ is the parity operator in the $(1/2,0)\oplus(0,1/2)$ representation space~\cite{Speranca:2013hqa,Ahluwalia:2019etz}. 
The new dual gives the following orthonormality relations 
 \begin{equation}
\gdualn{\lambda}_i(\p) {\lambda}_j(\p) =  2m \delta_{i j},\quad i,j=1,2,3,4\label{eq:on1234a}
\end{equation}
The associated spin sums evaluate to
\begin{align}
\sum_{i=1,2}{\lambda}_i(\p) \gdualn{\lambda}_i(\p) & = 
\left(
\begin{array}{cccc}
 m & 0 & i (\text{p0}+\text{pz}) & i \text{px}+\text{py} \\
 0 & m & i \text{px}-\text{py} & i (\text{p0}-\text{pz}) \\
 -i (\text{p0}-\text{pz}) & i \text{px}+\text{py} & m & 0 \\
 i \text{px}-\text{py} & -i (\text{p0}+\text{pz}) & 0 & m \\
\end{array}\right)\label{eq:ss12}
\end{align}
and 
\begin{align}
\sum_{i=3,4}{\lambda}_i(\p) \gdualn{\lambda}_i(\p) & = 
\left(
\begin{array}{cccc}
 m & 0 & -i (\text{p0}+\text{pz}) & -i \text{px}-\text{py} \\
 0 & m & \text{py}-i \text{px} & -i (\text{p0}-\text{pz}) \\
 i (\text{p0}-\text{pz}) & -i \text{px}-\text{py} & m & 0 \\
 \text{py}-i \text{px} & i (\text{p0}+\text{pz}) & 0 & m \\
\end{array}
\right)
 \label{eq:ss34}
\end{align}
Next, we define $a_\mu$ as coefficients of  $p^\mu 
= \{p^0,\p\} =
\{p^0,p_x,p_y,p_z\}$ in the above spins sums. The $a_\mu$ thus extracted read
\begin{align}
& a_0 = \left(
\begin{array}{cccc}
 0 & 0 & i & 0 \\
 0 & 0 & 0 & i \\
 -i & 0 & 0 & 0 \\
 0 & -i & 0 & 0 \\
\end{array}
\right),\quad
a_1=\left(
\begin{array}{cccc}
 0 & 0 & 0 & i \\
 0 & 0 & i & 0 \\
 0 & i & 0 & 0 \\
 i & 0 & 0 & 0 \\
\end{array}
\right)\\
& a_2=\left(
\begin{array}{cccc}
 0 & 0 & 0 & 1 \\
 0 & 0 & -1 & 0 \\
 0 & 1 & 0 & 0 \\
 -1 & 0 & 0 & 0 \\
\end{array}
\right),\quad
 a_3=\left(
\begin{array}{cccc}
 0 & 0 & i & 0 \\
 0 & 0 & 0 & -i \\
 i & 0 & 0 & 0 \\
 0 & -i & 0 & 0 \\
\end{array}
\right)
\end{align}
The spin sums then take the compact form
\begin{align}
\sum_{i=1,2}{\lambda}_i(\p) \gdualn{\lambda}_i(\p) & =
  a_\mu p^\mu + m \I
\label{eq:ss12a}\\
\sum_{i=3,4}{\lambda}_i(\p) \gdualn{\lambda}_i(\p) & = 
 - a_\mu p^\mu + m \I
 \label{eq:ss34b}
\end{align}
where  $\I$ represents the $4\times 4$ identity matrix. 
\vspace{11pt}

It is now in order that we explicitly draw attention of our reader to the facts, (a) that all the norms in equation (\ref{eq:on1234a}) are positive definite, (b) that in contrast to the spin sums of the Dirac formalism, the spin sums here carry same sign for the mass term, and opposite for the $a_\mu p^\mu$ terms, and (c) the completeness relation following from the spin sums carries a plus (between `1-2' and `3-4' parts), rather than, a minus sign of the Dirac counterpart:

\begin{equation}
\frac{1}{2 m}
\left[\sum_{i=1,2} {\lambda}_i(\p) \gdualn{\lambda}_i(\p) + 
\sum_{i=3,4} {\lambda}_i(\p) \gdualn{\lambda}_i(\p) \right]= \I .\label{eq:spinsums}
\end{equation}
These differences are crucial for the bosonic quantum field under construction.

For a direct comparison of $a_\mu $ with $\gamma_\mu$ we introduce $\s = (\sigma_x,\sigma_y,\sigma_z)$, where the $\sigma$'s are the Pauli matrices in the standard representation. Then the $a_\mu$ and  $\gamma_\mu$ may be written as (reminder: both in the Weyl representation)
\begin{align}
a_0 &= i \left(\begin{array}{cc}
\0 & \I\\
-\I & \0
\end{array}\right),\quad
\a  = i \left(\begin{array}{cc}
\0 &\s\\
\s & \0
\end{array}\right),\quad
\gamma_0 &= 
\left(\begin{array}{cc}
\0 & \I\\
\I & \0
\end{array}\right),\quad
\gb  =  \left(\begin{array}{cc}
\0 &\s\\
- \s & \0
\end{array}\right)
\end{align}
Using Pauli matrices $\s$, we may define $\sigma_\mu=(\sigma_0,\s)$, and $\overline{\sigma}_\mu=(\overline{\sigma}_0,\overline{\s})$ with $\sigma_0 = \overline{\sigma}_0$  identified with $2\times 2$ identity matrix $\I$, and $\overline{\sigma} = - \s$. Then the $a_\mu$ and $\gamma_\mu$ may be written as
\begin{equation}
a_\mu = i \left(
\begin{array}{cc}
\0 & \sigma_\mu\\
-\overline{\sigma}_\mu &\0
\end{array}
\right), \quad \gamma_\mu=
\left(\begin{array}{cc}
\0 & \sigma_\mu\\
\overline{\sigma}_\mu & \0
\end{array}
\right)
\end{equation}
With $\gamma$ and $a$ defined in (\ref{eq:gamma5}) and  
 (\ref{eq:gamma5d}) below, it follows that
\begin{equation}
a_\mu = i \gamma\, \gamma_\mu, \quad
\gamma_\mu = i a \, a_\mu
\end{equation}

\section{A bosonic representation of Clifford algebra}\label{Sec:CliffordAlgebra}

The $a_\mu$ may be compared with the $\gamma_\mu$.  Despite the manifest difference, they satisfy the Clifford algebra
\begin{equation}
\{a_\mu, a_\nu\} =
2\eta_{\mu\nu}\label{eq:ca}
\end{equation}
Furthermore, $a_\mu$ and $\gamma_\mu$ do not commute 
\begin{equation}
\left[ a_\mu,\gamma_\nu \right] = 2   \eta_{\mu\nu}\, i \gamma
\end{equation}
where $\gamma$ is defined as
\begin{equation}
 \gamma = \frac{i}{4!} \epsilon^{\mu\nu\lambda\sigma}
\gamma_\mu\gamma_\nu\gamma_\lambda\gamma_\sigma = \left(\begin{array}{cc}
\I_2 &\0_2\\
\0_2 & -\I_2
\end{array}\right)\label{eq:gamma5}
\end{equation}
with  $\epsilon^{\mu\nu\lambda\sigma}$ the
completely antisymmetric 4th rank tensor with $\epsilon^{0123} = +1$. The  $\0_2$ and $\I_2$ are  $2\times 2$ null and identity matrices, respectively. Similarly, we define
\begin{equation}
a = \frac{i}{4!}\epsilon^{\mu\nu\lambda\sigma} a_\mu a_\nu a_\lambda a_\sigma = - \left(\begin{array}{cc}
 \I_2 &\0_2\\
\0_2 & - \I_2
\end{array}\right)\label{eq:gamma5d}
\end{equation} 

The above-obtained spin sums when coupled with the orthonormality relations 
 (\ref{eq:on1234a})  give the identities
 \begin{align}
 \left(a_\mu p^\mu - m \I \right)\lambda_i(\p) = 0, \qquad \mbox{for $i=1,2$}\label{eq:xx}\\
 \left(a_\mu p^\mu + m \I \right)\lambda_i(\p) = 0. \qquad \mbox{for $i=3,4$}\label{eq:yy}
 \end{align}
The linear superpositions (\ref{eq:l12}) and (\ref{eq:l34}) thus make the $\lambda_i(\p)$ eigenspinors of $a_\mu p^\mu$. Since the $a_\mu$ -- as seen in the next section -- are embedded in a bosonic field theoretic structure we may consider the representation that we have obtained as a bosonic representation of the Clifford algebra -- in the sense just stated.

The results~(\ref{eq:ca}), (\ref{eq:xx}) and~(\ref{eq:yy}) are consistent with early work of Brauer and Weyl's~\cite{BrauerWeyl:1935}. 

\section{A quantum field with $\lambda_i(\p)$ as its expansion coefficients}\label{Sec:quantumfield}

Having developed the expansion coefficients and the associated formalism, we now explore the properties of the quantum field

\begin{equation}
\mathfrak{b}(x) \stackrel{\mathrm{def}}{=}
\int\frac{\mbox{d}^3 p}{(2\pi)^3}
\frac{1}{\sqrt{2 p^0}}
\bigg[
\sum_{i=1,2} {a}_i(\p)\lambda_i(\p) e^{-i p\cdot x}+ \sum_{i=3,4} b^\dagger_i(\p)\lambda_i(\p) e^{i p\cdot x}\bigg]\label{eq:fieldb}
\end{equation}
with its adjoint defined as
\begin{equation}
\gdualn{\mathfrak{b}}(x) \stackrel{\mathrm{def}}{=}
\int\frac{\mbox{d}^3 p}{(2\pi)^3}
\frac{1}{\sqrt{2 p^0}}
\bigg[
\sum_{i=1,2} a^\dagger_i(\p)\gdualn{\lambda}_i(\p) e^{i p\cdot x}+ \sum_{i=3,4} b_i(\p)\gdualn{\lambda}_i(\p) e^{-i p\cdot x}\bigg]
\end{equation}
At this stage we do not fix the statistics to be fermionic 
\begin{equation}
\left\{a_i(\p),a^\dagger_j(\p)\right\} =(2 \pi)^3 \delta^3(\p-\p^\prime)\delta_{ij}, \quad \left\{a_i(\p), a_j(\p^\prime)\right\} = 0 =
 \left\{a^\dagger_i(\p), a^\dagger_j(\p^\prime)\right\}\label{eq:b}
\end{equation}
or bosonic
\begin{equation}
\left[a_i(\p),a^\dagger_j(\p)\right] =(2 \pi)^3 \delta^3(\p-\p^\prime)\delta_{ij}, \quad \left[a_i(\p), a_j(\p^\prime)\right] = 0 =
 \left[a^\dagger_i(\p), a^\dagger_j(\p^\prime)\right]\label{eq:bd}
\end{equation} and assume similar anti-commutation, or commutation, relations for $b_i(\p)$ and $b_i^\dagger(\p)$. 


\subsection{The new field is bosonic}\label{sec:bosonic}

Following Weinberg~\cite[Page 198]{Weinberg:1995mt}, we note that for spacelike seprations Lorenz invariance of the $S$-matrix requires the field and its adjoint to commute (for bosons) and anticommute (for fermions).

Using the above obtained spin sums we find that 
\begin{align}
\left[\mathfrak{b}(x),\gdualn{\mathfrak{b}}(x^\prime) \right]_\mp
= 
{\big(}
\left[ i a_\mu \partial^\mu + m \I \right] \Delta_+(x^\prime-x) 
\nonumber \\
\mp
\left[- i a_\mu \partial^\mu + m \I \right] \Delta_+(x-x^\prime) {\big)}\label{eq:bosonic}
\end{align}
where
\begin{equation}
\Delta_+(x-x^\prime)\equiv \int\frac{d^3 p}{  (2\pi)^3\, 2p_0   } e^{i p\cdot( x-x^\prime)}
\end{equation}
For $(x-x^\prime)$ spacelike,   $\Delta_+(x-x^\prime)$ is an even function of $(x-x^\prime)$, and its first derivative is an  odd function of $(x-x^\prime)$. With this observation the field must be bosonic. That is,  in~(\ref{eq:bosonic}) we must choose the minus sign in $\mp$. The possibility (\ref{eq:b}) should be discarded, and the option (\ref{eq:bd}) adopted.
Thus for $(x-x^\prime)$ spacelike we have
\begin{equation}
\left[\mathfrak{b}(x),\gdualn{\mathfrak{b}}(x^\prime) \right]
=  0
\end{equation}
while for $t=t^\prime$, reduces the result~(\ref{eq:bosonic}) to 
\begin{equation}
\left[\mathfrak{b}(\x,t),\gdualn{\mathfrak{b}}(\x^\prime,t) \right]
=   a_0 \delta^3(\x-\x^\prime) \label{eq:bbbar}
\end{equation}

\subsection{Feynman-Dyson propagator}\label{Sec:FDPropagator}

Adapting a calculation similar to the one presented in Sections 15.3 and 15.4 of~\cite{Ahluwalia:2019etz}, the amplitude for a bosonic particle to propagate from $x$ to $x^\prime$ is given by
\begin{align}
\mathcal{A}_{x\rightarrow x^\prime} &= \mathrm{Amp}(x\to x^\prime, \mbox{particle})\Big\vert_{ t^\prime > t}   + 
\mathrm{Amp}(x\to x^\prime, \mbox{antiparticle})\Big\vert_{ t >t^\prime } \nonumber\\
&= \xi\Big(\underbrace{
\langle~\vert\mathfrak{b}(x^\prime) \gdualn{\mathfrak{b}}(x) \vert ~\rangle\theta(t^\prime-t) +
\left\langle~\right\vert\gdualn{\mathfrak{b}}(x) \mathfrak{b}(x^\prime) \vert ~\rangle\theta(t^\prime-t)}_{\langle~\vert     \mathfrak{T} [
\mathfrak{b}(x^\prime)\gdualn{\mathfrak{b}}(x)]\vert~\rangle
  }
\Big) \label{eq:Ampxxprime}
\end{align}
where $\langle~\vert\cdots\vert~\rangle$ represents vacuum expectation value of $\cdots$, and $\xi\in\mathbb{C}$  is determined below. The $\mathfrak{T}$ is the time ordering operator. In calculating $\mathcal{A}_{x\rightarrow x^\prime} $ the spin sums (\ref{eq:ss12}) and (\ref{eq:ss34}) play a crucial role.




The two vacuum expectation values that appear in $\mathcal{A}_{x\to x^\prime} $ evaluate to the following expressions
 \begin{align}
\langle\hspace{3pt}\vert
\mathfrak{b}(x^\prime)\gdualn{\mathfrak{b}}(x)\vert\hspace{3pt}\rangle  & =\int\frac{\text{d}^3p}{(2 \pi)^3}\left(\frac{1}{2  p^0}\right)
 e^{-ip\cdot(x^\prime-x)}
 \sum_{i=1,2}\lambda_i(\p)
\gdualn\lambda_i(\p) \nonumber \\
&=\int\frac{\text{d}^3p}{(2 \pi)^3}\left(\frac{1}{2  p^0}\right)
 e^{-ip\cdot(x^\prime-x)}
\left(a_\mu p^\mu + m\I\right)
\label{eq:amplitudeP-newS}
\end{align}
and 
\begin{align}
\langle\hspace{3pt}\vert
\gdualn{\mathfrak{b}}(x) \mathfrak{b}(x^\prime)\vert\hspace{3pt}\rangle 
&= \int\frac{\text{d}^3p}{(2 \pi)^3}\left(\frac{1}{2  p^0}\right)
 e^{- ip\cdot(x- x^\prime)}  \sum_{i=3,4}\lambda_i(\p)
\gdualn\lambda_i(\p)\nonumber\\
& = \int\frac{\text{d}^3p}{(2 \pi)^3}\left(\frac{1}{2  p^0}\right)
 e^{- ip\cdot(x- x^\prime)}
\left[- \left(a_\mu p^\mu - m\I\right) \right]\label{eq:amplitudeP-newA}
\end{align}
The above structures are exactly the same as for the Dirac case with two crucial differences,
(a) in equation (\ref{eq:Ampxxprime}) there is a plus, rather than a minus sign, between the two vacuum expectation values, and 
(b) in equation (\ref{eq:amplitudeP-newA}) there is a minus sign -- as compared with its Dirac counterpart -- in the spin sum written between the square brackets.

For the  step functions in equation
(\ref{eq:Ampxxprime}) we can use the Fourier representation
\begin{align}
\theta(t) = \frac{-1}{2\pi i}\int_{-\infty}^{\infty}
\frac{\exp(-i s t)}{s +i \epsilon}\mathrm{d}s
\end{align}
with $\epsilon,s\in\R$, to arrive at  (see, Weinberg's arguments in~\cite[Section 6.2]{Weinberg:1995mt}) 
\begin{equation}
\mathcal{A}_{x\to x^\prime}  =  i  \xi \int\frac{\text{d}^4 q}{(2 \pi)^4}\,
e^{-i q_\mu(x^{\prime\mu}-x^\mu)}
\frac{a_\mu q^\mu + m\I}{q_\mu q^\mu -m^2 + i\epsilon}\label{eq:AmplitudeWithXi}
\end{equation}
with $q^0$ no longer restricted to the on shell value $\sqrt{\q^2+m^2}$.
We immediately see that
\begin{equation}
\left( i a_{\mu^\prime}    \partial^{\mu^\prime}  - m \I \right)
\mathcal{A}_{x\to x^\prime}   =  i \xi\delta^4(x^\prime-x)
\end{equation}
The choice $\xi =  i$ makes $\mathcal{A}_{x\to x^\prime}$ the Green's function for the differential operator 
$\left( i a_{\mu^\prime}    \partial^{\mu^\prime}  - m \I \right)$ with the boundary condition specified by the $i\epsilon$ in the denominator. We thus define the Feynman-Dyson fropagator for the theory to be
\begin{equation}
\Delta_{FD}(x,x^\prime)=  i \mathcal{A}_{x\to x^\prime}  
\end{equation}
As a consequence  the field $\mathfrak{b}(x)$ is assigned a mass 
dimensionality  of three half~(see,~\cite[Section 12.1]{Weinberg:1995mt})
\begin{equation}
\mathcal{D}_\mathfrak{b} = \frac{3}{2}
\end{equation}
with the Lagrangian density
\begin{equation}
\mathfrak{L}(x) =
\gdualn{\mathfrak{b}} (x)\left(i a_\mu \partial^\mu - m \I\right) 
\mathfrak{b}(x)
\end{equation}
\subsection{Locality}\label{Sec:locality}

The momentum conjugate to $\mathfrak{b}(x)$ is
\begin{equation}
\pii(x)   = \frac{\partial \mathfrak{L}(x)}
{\partial {\dot{\mathfrak{b}}(x)}}  = i \gdualn{\mathfrak{b}} (x) a_0
\end{equation}
Combined with (\ref{eq:bbbar}) it gives
\begin{equation}
\left[\mathfrak{b}(\x,t),\;\pii(\x^\prime,t) \right] = i \delta^3\left(\x-\x^\prime\right)\I\quad\label{eq:lac-2and3c}
\end{equation}
A straight forward calculation adds to this, the following
\begin{equation}
\left[ \mathfrak{b}(\x,t),\;\mathfrak{b}(\x^\prime,t) \right]= 0, \quad 
\left[ \pii(\x,t),\;\pi(\x^\prime,t) \right] = 0,\label{eq:lac-2and3}
\end{equation}
establishing the field to be local.
\subsection{Field Energy: bounded from below}\label{Sec:FieldEnergy}

To examine if the energy associated with the introduced $\mathfrak{b}(x)$-$\gdualn{\mathfrak{b}}(x)$ system has  the usual zero point contribution and is bounded from below, we carry out a calculation similar to the one presented in \cite[Section 7]{Ahluwalia:2004ab} and find the field energy to be
\begin{equation}
H = \int\frac{\text{d}^3 p}{(2\pi)^3}\frac{1}{2 m} E(\p)\Bigg[\sum_{i=1,2} a_i^\dagger(\p) a_i(\p) \gdualn{\lambda}_i(\p) \lambda_i(\p) 
+
\sum_{i=3,4} b_i(\p) b_i^\dagger(\p) \gdualn{\lambda}_i(\p) \lambda_i(\p)
\Bigg]
\end{equation}
Use of the orthonormality relations~(\ref{eq:on1234a}) reduce the above expression to
\begin{equation}
H = \int\frac{\text{d}^3 p}{(2\pi)^3} E(\p)\Bigg[\sum_{i=1,2} a_i^\dagger(\p) a_i(\p) 
+
\sum_{i=3,4} b_i(\p) b^\dagger_i(\p)
\Bigg]
\end{equation}
The next simplification occurs by exploiting
\begin{equation}
\left[ b_i(\p),b_i(\p^\prime)\right]=(2\pi)^3\delta^3(\p-\p^\prime)\label{eq:stat}
\end{equation}
with the result that
\begin{equation}
H= \underbrace{\delta^3(0) \int\text{d}^3p \; 2 E(\p) }_{H_0}\;+ \sum_{i=1,2} \int\frac{\text{d}^3 p}{(2 \pi)^3}
E(\p) a_i^\dagger(\p) a_i(\p)
+ \sum_{i=3,4} \int\frac{\text{d}^3 p}{(2 \pi)^3}
E(\p) b_i^\dagger(\p) b_i(\p)\nonumber
\end{equation}
To obtain a representation for $\delta^3(0) $ that appears in the above expression for the field energy, we note that since $\delta^3(\p)$ may be expanded as 
\begin{equation}
\delta^3(\p) = \frac{1}{(2\pi)^3}\int\text{d}^3 x \exp(i \p\cdot \x)
\end{equation}
 $\delta^3(0)$ may be replaced by  $[1/(2\pi)^3]\int\text{d}^3 x$, giving the following contribution for the zero point energy
 \begin{equation}
 H_0 =  4 \times \frac{1}{(2\pi)^3}\int\text{d}^3 x\int\text{d}^3 p \; \frac{1}{2} E(\p)
 \end{equation}
 Since in natural units $\hbar$ is set to unity, $ \frac{1}{(2\pi)^3} \text{d}^3 x \,\text{d}^3 p $ acquires the interpretation of a unit-size phase cell,  with $\frac{1}{2} E(p)$ as its energy content.
The factor of $4$ in the expression for $H_0$ has the interpretation of four particle and antiparticle degrees carried
by the $\mathfrak{b}(x)$-$\gdualn{\mathfrak{b}}(x)$ system. The remaining two terms in the expression for $H$ establish that for a given momentum $\p$ each of the four particle-antiparticle degrees of freedom contributes equally.

\section{Concluding Remarks}\label{Sec:ConcludingRemarks}

\hfill  It has always been somewhat of a miracle that 
 Bosons \\  
 \vspace{-17pt}
 
 \hfill and Fermions, ``live such independent lives''.

\hfill Max Dresden in \cite[page 302]{Dresden:1985rs}

No longer so! This apparent miracle  has here been transformed into a magic: the $(1/2,0)\oplus(0,1/2)$ representation space that provides home to all the fermions of the standard model of high energy physics (SM), also provides a home at equal status to bosons. These can only be produced in pairs, a circumstance that could in the absence of the unexpected theoretical discovery presented here could easily go un-noticed or worst misinterpreted. Unlike a similarly surprising theoretical discovery of mass
dimension one fermions, the new bosons are of mass dimension three half. They can, therefore, extend SM doublets and be invoked in new theories of supersymmetry. At currently available accelerators
\begin{itemize}
\item if the new bosons appear together with the SM fermions in the interaction Lagrangian, then all such interactions are suppressed by at least two powers of unification scale
\item  on the other hand, if the new bosons appear together with the new fermions of~\cite{Ahluwalia:2019etz}, then all such interactions are suppressed by at least one power of unification scale
\end{itemize}
In early universe, the new bosons and fermions slosh back and forth with equal ease. The physical significance of the new bosons in cosmic structure formation, and later in star and galaxy formation, cannot but hold unsuspected surprises. 

The new bosons have the same number of degree of freedom, that is four, as those of SM fermions, or fermions of mass dimension one. Pairwise, the new bosons and fermions contribute zero to the cosmological constant. 

Supersymmetry extended the algebra of spacetime and unified fermions and bosons. From our perspective, its beginnings lie in a paper of Volkov and Akulov
written under the title~\cite{Akulov:1974xz}, ``Goldstone fields with spin $1/2$." Here, without extending the spacetime algebra we bring a unification of a different sort where fermions and bosons reside in the same representation space. \vspace{21pt}


\vspace{21pt}

\noindent
\textbf{Funding.} The research presented here is entirely supported by the personal funds of the author.
\vspace{5pt}

\noindent
\textbf{Competing interests.} The author declares no competing interests.

\noindent
\textbf{Acknowledgements.} I am grateful to the two anonymous referees who reviewed and commented constructively. I thank  
Julio Marny Hoff da Silva and Cheng-Yang Lee for correspondence related to the ideas presented here, and Sweta Sarmah for discussions.

\newpage

\newpage
\noindent



\appendix

\section{CPT for $\lambda(\p)$ spinors and Table A.1}

For a reader who may wish to study the CPT properties
of $\mathfrak{b}(x)$, the properties of
$\lambda(\p)$ under various discrete symmetries may be helpful. 
Towards that task, introduce $\Theta$, the Wigner time reversal operator for spin one half
\begin{align}
&\Theta =\left(\begin{array}{cc}
0 &-1\\
1 &0
\end{array}
\right)\end{align}
and recall the definition of $\gamma$ from~(\ref{eq:gamma5})
The charge conjugation $\mathcal{C}$, parity $\mathcal{P}$, and time reversal $\mathcal{T}$, operators can then be written as
\begin{equation}
\mathcal{C} = \left(\begin{array}{cc}
\0 & i\Theta \\
-i\Theta &\0
\end{array}
\right) K,\quad
\mathcal{P}= m^{-1} \gamma_\mu p^\mu,\quad
\mathcal{T} = i \gamma \mathcal{C}
\end{equation}
where $K$ complex conjugates to the right. With these definitions  we obtain
\begin{align}
&\mathcal{C} \lambda_1(\p) = i \lambda_4(\p),\quad
\mathcal{C} \lambda_2(\p) = -  i  \lambda_3(\p),\quad
\mathcal{C} \lambda_3(\p) =  - i\lambda_2(\p),\quad
\mathcal{C} \lambda_4(\p) =  i \lambda_1(\p), \label{eq:C}\\
&\mathcal{P} \lambda_1(\p) = i \lambda_3(\p),\quad
\mathcal{P} \lambda_2(\p) =  i \lambda_4(\p),\quad
\mathcal{P} \lambda_3(\p) = - i \lambda_1(\p),\quad
\mathcal{P} \lambda_4(\p) =  - i \lambda_2(\p), \label{eq:P}\\
&\mathcal{T} \lambda_1(\p) =   \lambda_2(\p),\quad
\mathcal{T} \lambda_2(\p) =   -  \lambda_1(\p),\quad
\mathcal{T} \lambda_3(\p) =  -\lambda_4(\p),\quad
\mathcal{T} \lambda_4(\p) =  \lambda_3(\p)\label{eq:T}
\end{align}
with the consequence that $(\mathcal{CPT})^2=\I$, with $\mathcal{C}^2=\I$, $ \mathcal{P}^2=\I$, $\mathcal{T}^2=-\I$. The charge conjugation and parity operators anticommute:
$
\{\mathcal{C},\mathcal{P}\} = 0.
$

\begin{table}\label{tab:trinity}
\begin{minipage}{485pt}
\caption{The plus sign indicates that $\Omega_\ell$ and 
$\Omega_{\ell'}$ anti-commute, while the minus sign indicates that they commute. For example, $\left[\Omega_7,\Omega_3\right] =0$ while 
$\left\{\Omega_{16},\Omega_{15}\right\}=0$. In the table below
the plus and minus sign appear in equal numbers. Each row, and each colum has eight plus and eight minus signs.
}\label{table1}
\addtolength\tabcolsep{1pt}\vspace{5pt}
\begin{tabular}{@{}ccccccccccccccc@{\hspace{20pt}}ccc@{\hspace{10pt}}}
\hline\hline
\empty & $\Omega_1$ & $\Omega_2$ & $\Omega_3$ & $\Omega_4$ & $\Omega_5$ & $\Omega_6$
& $\Omega_7$ & $\Omega_8$ & $\Omega_9$
& $\Omega_{10}$ & $\Omega_{11}$ & $\Omega_{12}$
& $\Omega_{13}$ & $\Omega_{14}$ & $\Omega_{15}$ 
& $\Omega_{16}$ 
\\
\hline
$\Omega_1$ & $-$ & $-$  & $-$ 
& $-$  & $-$  & $-$ & $-$  & $-$  & $-$ & $-$  & $-$  & $-$ & $-$  & $-$  &$-$  &$-$  \\ 
$\Omega_2$ & $-$  & $-$  & $+$
& $+$  & $+$ & $-$& $+$ & $+$ & $+$& $-$ & $-$ & $+$& $-$ & $-$ & $-$ &$+$ \\ 
$\Omega_3$ & $-$ & $+$ & $-$
& $+$ & $+$& $+$& $-$& $+$ & $-$ & $+$& $-$ & $-$ & $+$& $-$ & $-$ &$+$ \\ 
$\Omega_4$ & $-$ & $+$ & $+$
& $-$ & $+$ & $+$& $+$ & $-$ & $-$& $-$ & $+$ & $-$& $-$ & $+$ 
&$-$ &$+$ \\ 
$\Omega_5$ & $-$ & $+$ & $+$
& $+$ & $-$ & $-$& $-$ & $-$ & $+$& $+$ & $+$ & $-$& $-$ & $-$ &$+$ &$+$ \\ 
$\Omega_6$ & $-$ & $-$ & $+$
& $+$ & $-$ & $-$& $+$ & $+$ & $-$& $+$ & $+$ & $-$& $+$ & $+$ &$-$ &$-$ \\ 
$\Omega_7$ & $-$ & $+$ & $-$
& $+$ & $-$ & $+$& $-$ & $+$ & $+$& $-$ & $+$ & $+$& $-$ & $+$ &$-$ &$-$ \\ 
$\Omega_8$ & $-$ & $+$ & $+$
& $-$ & $-$ & $+$& $+$ & $-$ & $+$& $+$ & $-$ & $+$& $+$ & $-$ &$-$ &$-$ \\ 
$\Omega_9$ & $-$& $+$ & $-$
& $-$ & $+$ & $-$& $+$ & $+$ & $-$& $+$ & $+$ & $+$& $-$ & $-$ & $+$ &$-$ \\ 
$\Omega_{10}$ & $-$ & $-$ & $+$
& $-$ & $+$ & $+$& $-$ & $+$ & $+$& $-$ & $+$ & $-$& $+$ & $-$ &$+$ &$-$ \\ 
$\Omega_{11}$ &$ -$ & $-$ & $-$
& $+$ & $+$ & $+$& $+$ & $-$ & $+$& $+$ & $-$ & $-$& $-$ & $+$ &$+$ &$-$ \\ 
$\Omega_{12}$ & $-$ & $+$ & $-$
& $-$ & $-$ & $-$& $+$ & $+$ & $+$& $-$ & $-$ & $-$& $+$ & $+$ &$+$ &$+$ \\ 
$\Omega_{13}$ & $-$ & $-$ & $+$
& $-$ & $-$ & $+$& $-$ & $+$ & $-$& $+$ & $-$ & $+$& $-$ & $+$ &$+$ &$+$ \\ 
$\Omega_{14}$ & $-$ & $-$ & $-$
& $+$ & $-$ & $+$& $+$ & $-$ & $-$& $-$ & $+$ & $+$& $+$ & $-$ &$+$&$+$ \\ 
$\Omega_{15}$ & $-$ & $-$ & $-$
& $-$ & $+$ & $-$& $-$ & $-$ & $+$& $+$ & $+$ & $+$& $+$ & $+$ &$-$ &$+$ \\ 
$\Omega_{16}$ & $-$ & $+$ & $+$
& $+$ & $+$ & $-$& $-$ & $-$ & $-$& $-$ & $-$ & $+$& $+$ & $+$ &$+$ &$-$ \\ 
\hline\hline
\end{tabular} 
\end{minipage}
\end{table}

\newpage

\end{document}